\documentclass[prb,preprint]{revtex4}
\usepackage{epsfig,graphicx}
\begin{document}

\title{Sequential and co-tunneling behavior in the temperature-dependent
thermopower of few-electron quantum dots}

\author{R. Scheibner, E.G. Novik, T. Borzenko, M. K\"onig, D. Reuter$^*$, A.D.
Wieck$^*$, H. Buhmann, and L.W. Molenkamp}

\affiliation{Physikalisches Institut (EP3), Universit\"at
W\"urzburg, Am Hubland, 97074 W\"urzburg, Germany}

\affiliation{$^*$Lehrstuhl f{\"u}r Angewandte Festk{\"o}rperphysik,
Ruhr-Universit{\"a}t Bochum, Universit{\"a}tsstra{\ss}e 150, 44780 Bochum,
Germany}

\date{\today}

\begin{abstract}
We have studied the temperature dependent thermopower of
gate-defined, lateral quantum dots in the Coulomb blockade regime
using an electron heating technique. The line shape of the
thermopower oscillations depends strongly on the contributing
tunneling processes. Between 1.5~K and 40~mK a crossover from a
pure sawtooth- to an intermitted sawtooth-like line shape is
observed. The latter is attributed to the increasing dominance of
cotunneling processes in the Coulomb blockade regime at low
temperatures.
\\


\end{abstract}

\maketitle


The perspective of scalable semiconductor quantum processing
devices feeds an intense interest in quantum dot (QD) structures
that contain only a few electrons.\cite{Loss:quantumcomp} For the
development of these devices, a detailed knowledge of the
underlying electron transport processes is of crucial importance.
So far, most of the transport experiments have focused on the
electrical conductance.\cite{Kouwenhoven:overview} Although
thermoelectrical transport measurements are known to be more
sensitive to the details of the electronic structure than
conventional transport
measurements,\cite{ziman:electronsandphonons} little experimental
attention has been paid to this kind of measurements on QDs.
Within the scope of the Onsager relations, the thermopower $S$,
which is given by

\begin{equation}\label{sdefine}
S \equiv - \lim_{\Delta T \rightarrow 0} \left.\frac{V_{\rm
T}}{\Delta T}\right|_{I=0} =-\frac{\langle E \rangle}{eT}\,,
\end{equation}

relates the average energy $\langle E \rangle$ transfer at a
temperature $T$ to the thermovoltage $V_{\rm T}$ for a given
temperature difference $\Delta T$ across the device at zero net
current. This additional information about the carrier kinetics is
not provided by conventional transport measurements and helps to
distinguish between different possible transport
regimes.\cite{Scheibner:PRL:95:176602:2005}

In the past, thermopower measurements on QDs in the Coulomb
blockade (CB) regime have yielded qualitatively different results:
either a sawtooth-like line shape, or a line shape similar to the
derivative of a Coulomb Blockade (CB) conductance peak is
observed when the electrochemical potential is varied in order to
change the number of electrons occupying the QD. So far, a sawtooth-like
behavior has been observed mainly for many-electron QDs, while derivative-like
line shapes are predominantly reported for smaller QDs at low (mK)
temperatures.\cite{Staring:EPL:22:57:1993,S. Godjin,A.S.Dzurak}

Here, we present thermovoltage measurements on gate defined,
lateral QDs, containing a few tens of electrons, which allow us to
analyze the low temperature line shape profile in detail. For a
series of CB conductance resonances, the transition is observed from a full
sawtooth line shape to a sawtooth that is periodically intermitted
by a zero thermovoltage signal while the temperature
is lowered from $T = 1.5$~K to $T< 100$~mK. This behavior is in
qualitative agreement with recent theoretical considerations of
Turek and Matveev \cite{Turek:Matveev} for many-electron QDs. The
transition is associated with an increasing dominance of
cotunneling processes for decreasing temperatures. In our
measurements we find that the regime of sequential tunneling,
which dominates the transport in the vicinity of the CB
resonances, extends much further than anticipated for
many-electron dots. This leads to an enhanced absolute thermopower
for few-electron devices.


\begin{figure}
\centering
\includegraphics[width = 8.6 cm]{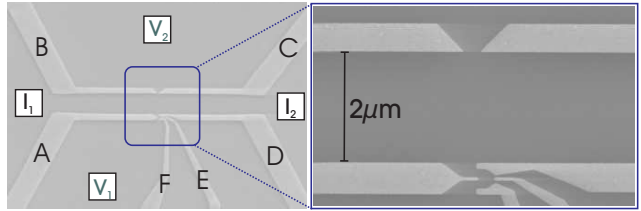}
\caption{Scanning electron microscope image of the sample
structure. Schottky-gates are labeled A, B, ... F. Sample areas
which serve as current and voltage contacts are labeled $I_{1}$,
$I_{2}$ and $V_{1}$, $V_{2}$, respectively.}\label{fig1}
\end{figure}

The measurements are carried out in a top-loading dilution
refrigerator at lattice temperatures between 40~mK and 1.5~K. The
GaAs/(Al,Ga)As QDs are fabricated by split-gate technology using
optical and electron-beam lithography. The two dimensional
electron gas (2DEG) is located 70 nm below the surface, has a
carrier density $n_e=2.3\times 10^{15} \text{~m}^{-2}$, and a
mobility $\mu = 100 \text{~m}^2/(\text{Vs})$. The gate structure
of the samples is shown in Fig.~\ref{fig1}.
\cite{m_ciorga_etal_PRB61nr24_2000_S16315} Gates A, D, E and F
form the QD with a nominal diameter of approximately $250$~nm. The
voltage applied to the plunger gate E, $V_{\rm E}$, is used to
control the number of electrons on the QD. Low-frequency lock-in
techniques are used for the transport experiments. The effective charging energy
$E^*_{\rm C}$ is determined From finite
bias conductance measurements. In the regime where CB oscillations
are observable, $E^*_{\rm C}$ decreases from 3.0~meV to 0.8~meV
when the coupling to the reservoirs is increased.

Gates B and C define, together with gates A and D, the electron
heating channel, and, at the same time, the reference quantum
point contact (QPC$_{\rm ref}$) in the thermopower
experiments.\cite{currentheating,qpcreference} By adjusting the
conductance of QPC$_{\rm ref}$ to the center of a quantized plateau,
its thermopower is approximately zero, $S_{\rm QPC_{ref}} \approx
0$. The transverse voltage $V_{\rm T}\equiv V_1-V_2$ (c.f.\ Fig.\
\ref{fig1}) is then directly proportional to the thermopower of
the QD. A constant temperature difference across the dot is
maintained by passing an ac current of $\nu$ = 13~Hz through the
heating channel. $V_{\rm T}$ is detected at twice the excitation
frequency (26~Hz). Here, we discuss two QD samples (QD1 and QD2) that have
the same gate structure design but adjusted to have a different number of electrons; QD1
(QD2) contains between 15 (30) and 20 (40) electrons. A current of
$I_{\rm H}=9.7$~nA (4.2 nA) increases the electron
temperature in the heating channel by $\Delta T$~=~9~mK (3 mK) for
QD1 (QD2).


\begin{figure}
\begin{centering}
\includegraphics[width = 8.6 cm]{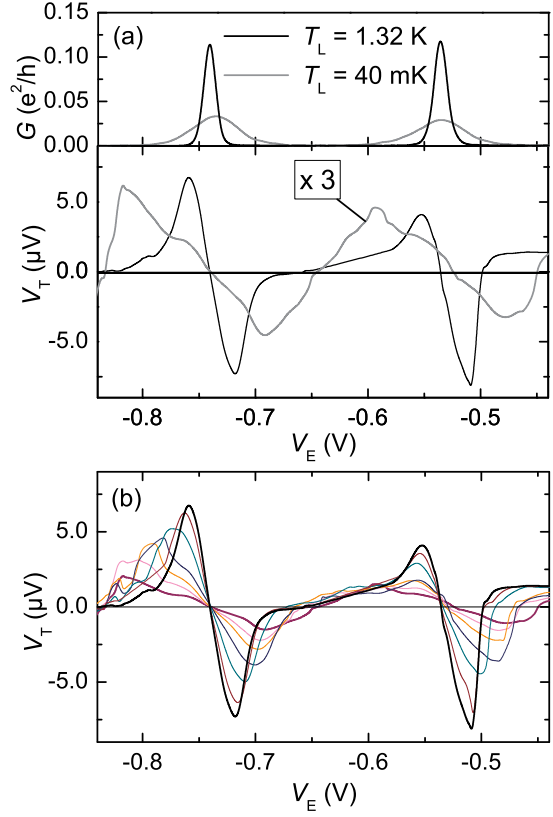}
\caption{(color online) (a) Conductance $G$ (upper panel) and the
corresponding thermovoltage $V_{\rm T}$ (lower panel) of QD1 as a
function of the plunger-gate voltage $V_{\rm E}$. (b)
Thermovoltage at seven different temperatures: $T_{\rm L}$ = 40 mK
(black), 66 mK (brown), 158 mK (cyan), 257 mK (blue), 425 mK
(orange), 1.04 K (pink) and 1.5 K (purple).}\label{fig2}
\end{centering}
\end{figure}

Figure \ref{fig2} (a) shows the conductance $G$ and the
corresponding thermovoltage $V_{\rm T}$ of QD1 for an
electrostatic charging energy $E_{\rm C} = 1.43$~meV at lattice
temperatures $T_{\rm L} = 40$~mK (black lines) and 1.32~K (grey
lines). At 1.32~K, thermal broadening determines the shape of the
CB resonances. The corresponding thermovoltage signal shows a
sawtooth-like line shape with a maximum in the vicinity of the
center of the CB. Additional fine structure on the sawtooth line shape
and is due to elevated temperatures and
finite level spacings in the QD. At $T_{\rm L}=40$~mK the
conduction resonance peaks have an increased height, a reduced
width, and are well separated by regimes of (approximately) zero
conductance. The line shape of the corresponding thermovoltage now
resembles
more the negative derivative of the conductance $G$. 
The values of the thermovoltage extrema increase by a factor of
three and their positions are shifted towards the CB resonances.

A small asymmetry between the thermovoltage values of positive and
negative amplitude is observed for all measurements. Lower tunnel
probabilities, for strongly negative plunger gate voltages
$V_{\rm E}$, reduce this asymmetry (c.f.\ Fig.\ \ref{fig2} and
Fig.\ \ref{fig4}). The asymmetry is intrinsic and has been
ascribed to multi channel tunneling processes \cite{XSC} and the
energy dependent transmission probability of the tunnel barriers.

Figure \ref{fig2}(b) shows $V_{\rm T}$ for seven different
temperatures in the range from $T_{\rm L}=$ 1.32~K down to 40~mK.
It is evident that the change in line shape occurs continuously.
In the vicinity of the CB resonance the thermovoltage varies
linearly with $V_{\rm E}$, its slope increases with decreasing $T$
while in-between the CB resonances, a region where $V_{\rm
T}\simeq 0$ develops. The observation of two different line shapes
indicates that at different temperatures different transport
mechanisms dominate the electronic transport properties.

Near the CB resonances, the charge transport is dominated by
sequential tunneling (ST) processes and is explained within the
so-called orthodox model\cite{Beenakker:tpseqtunneling}, where
only first order tunneling processes are considered. In-between
the resonances, the transport can also be due to ST processes, but
only at relatively high temperatures ($\sim 1 \;$K). In ST
transport, the average electron energy is proportional to an
effective energy gap $E_{\rm g}$ which is defined as the
difference between the Fermi energy of the leads and the energy of
the closest QD state. $E_{\rm g}$ varies linearly between
$-E_C^*/2$ and $+E_C^*/2$ with increasing electrochemical
potential of the QD, $\Phi_{\rm QD}$, and subsequently jumps back
to $-E_C^*/2$ at the center of the CB. According to
Eq.~(\ref{sdefine}), the thermovoltage follows this sawtooth-like
behavior. Thermal smearing at higher temperatures leads to a more
sinusoidal variation of $E_{\rm g}$ and thus the maxima of the
thermovoltage oscillations are located slightly away from the
center of the CB. The ST mechanism thus explains the line shape at
temperatures around 1 K.

We attribute the low (mK) temperature line shape to the occurrence
of (inelastic) cotunneling (CT) transport in between the CB
resonances.\cite{Turek:Matveev} At low temperatures, these higher
order processes dominate the transport away from the CB resonance,
because the ST processes are thermally activated and thus
exponentially suppressed on lowering the temperature, while CT
processes scale only according to a power
law.\cite{Averin:PRL:65:2446:1990} Due to energy conservation, the
average energy transferred by cotunneling processes is
proportional to the temperature ($\langle E_{\rm co}\rangle
\propto T$). Therefore, the expected thermoelectric signal of CT
processes is vanishingly small.\cite{Turek:Matveev} Decreasing the
sample temperature implies a transition from ST- to CT-dominated
transport in the CB regime away from the conductance resonances
and thus a suppression of the thermovoltage signal in the
corresponding gate voltage ranges. The sawtooth line shape becomes
intermitted by regions of (nearly) zero signal amplitude, as
observed in Fig.\ \ref{fig2}.

\begin{figure}
\centering
\includegraphics[width = 8.6 cm]{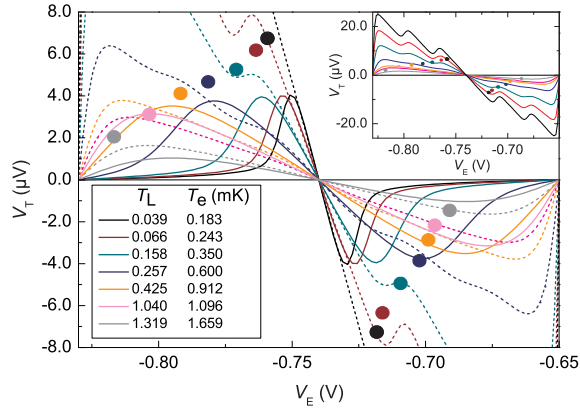}
\caption{(color online) Calculated thermovoltage for the orthodox
(dotted lines) and CT-included model (full lines) as a function of
$V_{\rm E}$. The dots indicate the maxima of the measured
thermovoltage signal. The inset show the orthodox model at full
scale.} \label{fig3}
\end{figure}

In order to discuss this transition more quantitatively we compare
the thermovoltage resonance at $V_{\rm E}=-0.73$ V with the
behavior of the orthodox (pure ST) \cite{Beenakker:tpseqtunneling}
model, and a model that also includes CT
effects\cite{Turek:Matveev}. At $T_{\rm L}$= 1.32 K both models
exhibit nearly the same thermovoltage amplitude and approximately
the same line shape [c.f. Fig.\ \ref{fig3}]. For this temperature,
our QD fulfills the condition $\hbar\Gamma<<k_BT<<E_{\rm C}$,
which allows us to extract the relevant model parameters. We
obtain $E_C^* = 1.712$~meV, $\alpha = \Phi_{\rm QD}/(-e V_{\rm
E})= 0.0095$, and $G_{l,r} = 0.072\, e2/h$, where $G_{l,r}$
describes the tunnel conductance to the left and right reservoir,
respectively. Considering these parameters as temperature
independent we calculate the temperature dependent thermovoltage
[Fig.~\ref{fig3}]. Since the charging energy of the QD determines
the slope of the thermovoltage signal in the close vicinity of the
CB resonances, the direct comparison with the experiment allows us
to extract the effective electron temperature $T_{\rm E}$ under
our experimental conditions. The deduced values for $T_{\rm E}$
are shown in Fig.~\ref{fig3} and have been confirmed independently
by fitting the conductance
peaks.\cite{Beenakker:gseqtunneling,Vorojtsov:2004:18:3915} For
clarity, only the maxima of the measured thermovoltage are
indicated in the figure, together with the results of the model
calculations.

In the orthodox model, a sawtooth line shape is predicted for
all temperatures (Fig.~\ref{fig3}). The wiggles at the declining
slope of the sawtooth come from excited states and have the
periodicity of the level spacing. At the same time, the
CT-included model does indeed reproduce a transition from a
sawtooth to a periodically suppressed sawtooth line shape.
However, while the CT-included model predicts an approximately
constant peak amplitude, the experiments show a strong increase in
peak amplitude with decreasing temperature. In addition, the model
does not predict the gate voltage position of the maxima
correctly. A comparison of Fig.~\ref{fig2} and Fig.~\ref{fig3}
reveals that in the experiment the linear increase of the
thermovoltage around the CB resonance extends much further than
anticipated by the CT-included model, and rather follows the
behavior of the orthodox model i.e.,
the voltage range where ST dominates the transport is larger than
given by the CT-included model. In other words, the CT-included
model overestimates the influence of CT processes for our
few-electron QD. The reason is that the CT-included model assumes a negligible energy
spacing of the QD states ($\delta E_{\rm QD} << k_BT$), as
applicable for metallic dots. For the present few-electron QD
$\delta E_{\rm QD}$ clearly is not negligible, the energy gap
between the ground and the first excited state is of the order of
250~$\mu e$V. Even at temperatures around 1 K, only a few excited
states are available for transport, which reduces the probability
for co-tunnel events.

For comparison, we measured a second QD sample, which has the same
layout as QD1 (c.f.\ Fig.\ \ref{fig1}). QD2 exhibits a similar
charging energy ($E^*_{\rm C} \approx$ 1.5 meV) but smaller level
spacing ($\delta E \sim 50$ $\mu$eV) presumably due to variations
in the potential landscape in the
2DEG.\cite{Koonen:PRL:84:2473:2000} Figure \ref{fig4} (a) shows
the change of the thermovoltage line shape for a series of CB
resonances. Clearly, this sample is further into the many electron
limit. The change in line shape with temperature is still present. It
becomes less pronounced for more positive gate voltages, where
the coupling of the QD states to the leads increases.
The line shape close to the CB resonance at $V_{\rm E}=-1.9$
V has been used to evaluate the model calculations similar to QD1.
Fig. \ref{fig4}
(b) confirms confirms that again the CT-included model
overestimates the influence of CT processes comparable to the case
of QD1.

\begin{figure} \centering
\includegraphics[width = 8.6 cm]{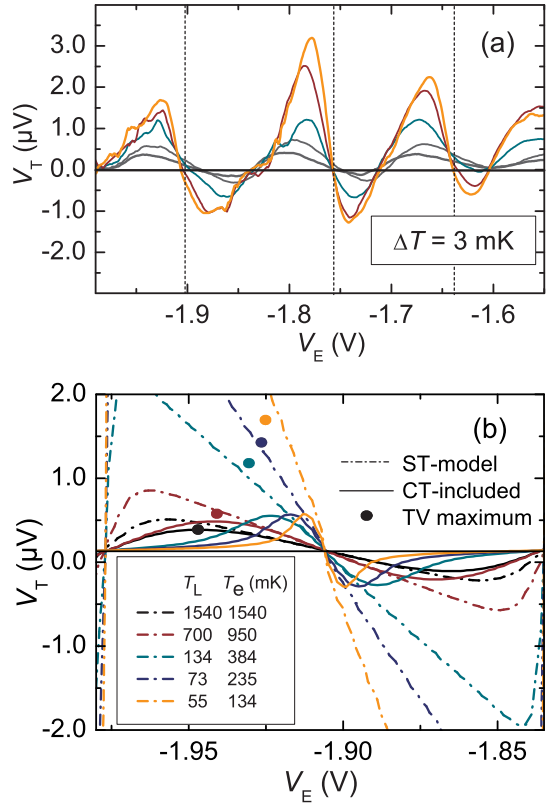}
\caption{(color online) (a) Thermovoltage for a series of CB
resonances as a function of the plunger gate voltage $V_{\rm E}$
for QD2. The positions of the CB resonances are indicated by
vertical dashed lines. (b) Calculated thermovoltage for the
orthodox (dash-dotted lines) and CT-included model (full lines) as
a function of $V_{\rm E}$. The dots indicate the maxima of the
measured thermovoltage signal near the resonance at $V_{\rm E} =
-1.9 V$. }\label{fig4}
\end{figure}


In summary, the thermopower of few-electron QDs reveals a
transition of the line shape as a function of the QD potential
from a full sawtooth to an intermitted sawtooth behavior. This
transition is directly related to a change in the dominant
transport processes. While close to the CB resonances, sequential
tunneling processes dominate, the full CB regime is dominated by
inelastic cotunneling processes. Compared with many-electron
(metallic) QDs, the regime of sequential tunneling is extended to
a wider gate voltage range for few-electron QDs, which results in
an increasing thermopower peak amplitude with decreasing
temperature.

We wish to thank C. Gould and J. Weis for valuable
discussions. We gratefully acknowledge the financial support of the DFG
(Mo771/5-2) and the Office of Naval Research (04PR03936-00).

\newpage

\end{document}